\documentclass[useAMS,usenatbib]{mn2e}
\usepackage{graphics}
\usepackage{epsf}
\usepackage{subfigure}
\def\ni{\noindent}

\begin{document}

\title[LMC Wesenheit Function]{The Linearity of the Wesenheit function for the Large Magellanic Cloud Cepheids}
\author[Ngeow \& Kanbur]{Chow-Choong Ngeow$^1$\thanks{E-mail: ngeow@nova.astro.umass.edu} and Shashi M. Kanbur$^{1,2}$ 
\\
$^1$Department of Astronomy, University of Massachusetts,         
 Amherst, MA 01003, USA
\\
$^2$Department of Physics, State University of New York at Oswego, 
 Oswego, NY 13126, USA
}

%\begin{document}

\date{Accepted 2005 month day. Received 2005 month day; in original form 2004 November 30}

%\pagerange{\pageref{1}--\pageref{15}} \pubyear{2003}

\maketitle

\begin{abstract}

  There is strong evidence that the period-luminosity (PL) relation for the Large Magellanic Cloud (LMC) Cepheids shows a break at a period around 10 days. Since the LMC PL relation is extensively used in distance scale studies, the non-linearity of the LMC PL relation may affect the results based on this LMC calibrated relation. In this paper we show that this problem can be remedied by using the Wesenheit function in obtaining Cepheid distances. This is because the Wesenheit function is linear although recent data suggests that the PL and the period-colour (PC) relations that make up the Wesenheit function are not. We test the linearity of the Wesenheit function and find strong evidence that the LMC Wesenheit function is indeed linear. This is because the non-linearity of the PL and PC relations cancel out when the Wesenheit function is constructed. We discuss this result in the context of distance scale applications. We also compare the distance moduli obtained from $\mu_0=\mu_V-R(\mu_V-\mu_I)$ (equivalent to Wesenheit functions) constructed with the linear and the broken LMC PL relations, and find that the typical difference in distance moduli is $\sim \pm0.03$mag. Hence, the broken LMC PL relation does not seriously affect current distance scale applications. We also discuss the random error calculated with equation $\mu_0=\mu_V-R(\mu_V-\mu_I)$, and show that there is a correlation term that exists from the calculation of the random error. The calculated random error will be larger if this correlation term is ignored.

\end{abstract}


%************************************
%  BEGINNING OF TEXT
%************************************

\section{Introduction}

     The Cepheid period-luminosity (PL) relation plays a major role in distance scale studies, which can ultimately be used to determine the Hubble constant. The calibrating PL relation currently used is based mainly on the Large Magellanic Cloud (LMC) Cepheids, as applied by the $H_0$ Key Project team \citep{fre01} as well as in other studies \citep[e.g., ][]{sah01,kan03}. The Cepheid PL relation has long been considered to be a linear function of $\log(P)$ within the range of $\log(P)\sim0.3$ to $\log(P)\sim2.0$, where $P$ is the pulsation period in days. 

%************************************
%  FIGURE 1
%************************************

     \begin{figure*}
       \vspace{0cm}
       \hbox{\hspace{0.2cm}\epsfxsize=8.5cm \epsfbox{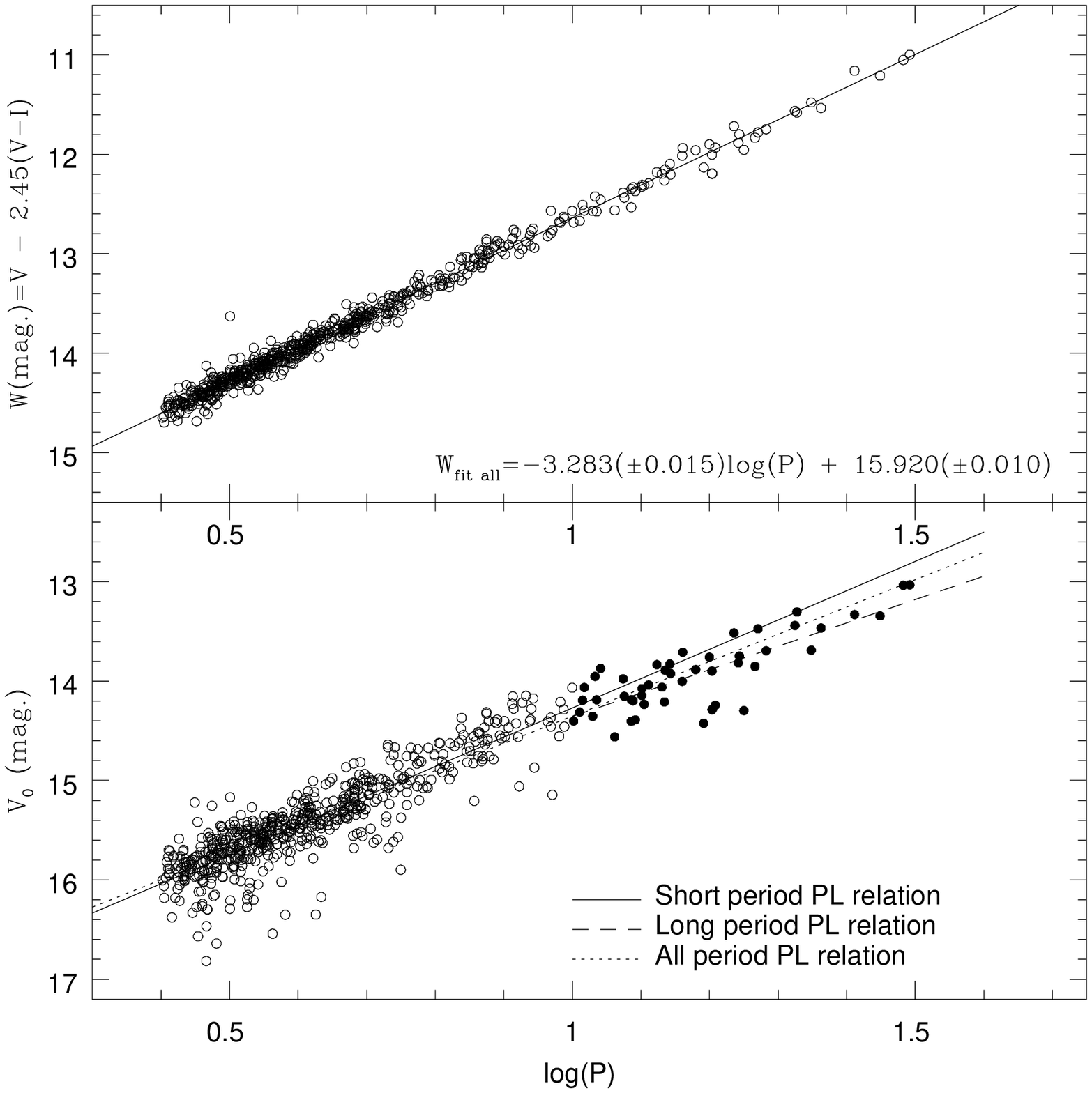}
         \epsfxsize=8.5cm \epsfbox{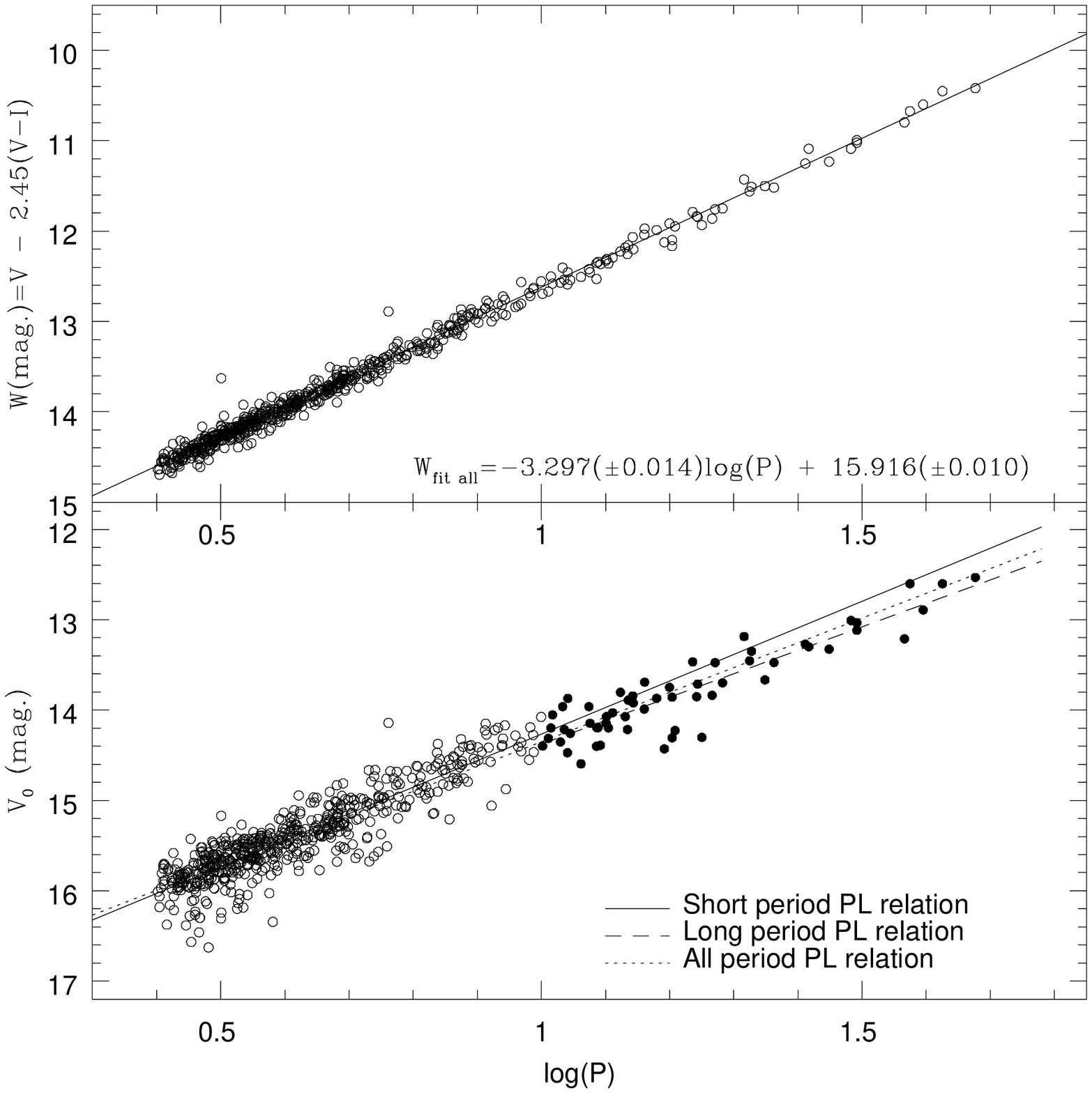}}
       \vspace{0cm}
       \caption{Comparison of the Wesenheit function with the extinction corrected {\it V}-band PL relation for the LMC Cepheids. On the top panels, we plot the Wesenheit function for the LMC data, which clearly show that the Wesenheit function is linear. We fit the data with a single line regression, using the least-squares method, as indicated on the figures. On the bottom panels, we show the extinction corrected {\it V}-band PL relations, and separate the data into the short (open circles) and long period (filled circles) Cepheids. The fitted PL relations for long, short and all period Cepheids are drawn as dashed, solid and dotted lines, respectively. We extend the short period PL relation to the longer period ranges in order to compare with the long period PL relations. The data in the left and right panels are taken from \citet{kan04} and \citet{kan05}, respectively.}
       \label{fig1}
     \end{figure*}

     However, the non-linearity of the LMC PL relation has been proposed by \citet{tam02a} and \citet{kan04}, i.e. the LMC data are more consistent with two PL relations and a discontinuity at a period around 10 days. This is illustrated in the lower panels of Figure \ref{fig1} for the extinction corrected {\it V}-band LMC PL relation. The existence of two LMC PL relations is further supported by the results from a rigorous statistical test (the $F$-test), as presented in \citet{kan04}, which shows that the {\it V}- and {\it I}-band LMC PL relations are better described by two PL relations. Since the work of \citet{tam02a} and \citet{kan04} are based on the OGLE (Optical Gravitational Lensing Experiment) LMC Cepheids \citep{uda99b}, which are truncated at $\log(P)\sim1.5$, \citet{san04} and \citet{kan05} used additional data that are available from the literature, especially those with $\log(P)>1.5$, to further support the existence of two PL relations in the LMC. These studies of the non-linear LMC PL relations are focused on the optical bands, as they are mainly based on the OGLE data, we hence discuss the non-linear LMC PL relation in the optical bands in this paper. The non-linear {\it V}-band LMC PL relation as seen from the OGLE data has been verified with the MACHO {\it V}-band data \citep{nge05}. In addition, \citet{nge05} also extended the study of the non-linear LMC PL relations to the {\it JHK}-bands with the 2MASS data. In Tables \ref{tab1} \& \ref{tab2}, we collect the slopes and the zero-points (ZP) for the long ($\log[P]>1.0$) and short period optical PL relations that are available in the literature\footnote{The {\it I}-band PL relations from \citet{kan04} are fitted with slightly different definition of the {\it I}-band mean magnitudes. Hence we refit the {\it I}-band PL relations with the conventional and reddening corrected {\it I}-band mean magnitudes. The results from the $F$-test remain unchanged.}, respectively. For completeness, we also include the linear, unbroken PL relation obtained using all Cepheids in these tables. The slopes and ZPs from different papers are consistent and agree with each other. Note that these LMC PL relations have been corrected for reddening. The reason that the LMC PL relation is non-linear is because the period-colour (PC) relation for LMC Cepheids is also non-linear across the 10 days period \citep{tam02a,kan04,san04,kan05}. The detailed investigation of the physics behind the broken LMC PL and PC relation is beyond the scope of this paper, but it is of great interest for the studies of stellar pulsation and evolution. 

%************************************
%  TABLE 1 & 2
%************************************

     \begin{table*}
       \centering
       \caption{Slopes for the LMC PL relations$^{\mathrm a}$. Long and short periods are referred to $P>10$ and $P<10$ days, respectively.}
       \label{tab1}
       \begin{tabular}{llcccc} \hline
         Band & Period-Range & TR02 & STR04 & KN04 & KNB05 \\        
         \hline \hline
         {\it V} & All   & $-2.760\pm0.031^{\mathrm b}$ & $-2.702\pm0.028$ & $-2.746\pm0.043$ & $-2.736\pm0.036$ \\
         {\it V} & Short & $-2.86\pm0.05$               & $-2.963\pm0.056$ & $-2.948\pm0.065$ & $-2.937\pm0.060$ \\
         {\it V} & Long  & $-2.48\pm0.17$               & $-2.567\pm0.102$ & $-2.350\pm0.252$ & $-2.598\pm0.161$ \\
         {\it I} & All   & $-2.962\pm0.021^{\mathrm b}$ & $-2.949\pm0.020$ & $-2.965\pm0.028$ & $-2.965\pm0.024$ \\
         {\it I} & Short & $-3.03\pm0.03$               & $-3.099\pm0.038$ & $-3.096\pm0.043$ & $-3.090\pm0.041$ \\
         {\it I} & Long  & $-2.82\pm0.13$               & $-2.822\pm0.084$ & $-2.737\pm0.179$ & $-2.918\pm0.112$ \\
         {\it B} & All   & $\cdots$                     & $-2.340\pm0.037$ & $\cdots$ & $\cdots$ \\
         {\it B} & Short & $-2.42\pm0.08$               & $-2.683\pm0.077$ & $\cdots$ & $\cdots$ \\
         {\it B} & Long  & $-1.89\pm0.62^{\mathrm c}$   & $-2.151\pm0.134$ & $\cdots$ & $\cdots$ \\
         \hline
       \end{tabular}
       \begin{list}{}{}
       \item   $^{\mathrm a}$ The references are: TR02 = \citet{tam02}; STR04 = \citet{san04}; KN04 = \citet{kan04}; KNB05 = \citet{kan05}
       \item   $^{\mathrm b}$ Since \citet{tam02} does not give the results from the fit to all Cepheids, we adopted the slopes from \citet{uda99a} because the same dataset is used in both papers.
       \item   $^{\mathrm c}$ The number of long period Cepheids in {\it B}-band is $\sim13$, hence the error is larger than the others. 
       \end{list} 
     \end{table*}
     
     \begin{table*}
       \centering
       \caption{Same as Table \ref{tab1}, but for the zero-points of the LMC PL relations, by assuming $\mu_{LMC}=18.50$mag.}
       \label{tab2}
       \begin{tabular}{llcccc} \hline
         Band & Period-Range & TR02 & STR04 & KN04 & KNB05 \\        
         \hline \hline
         {\it V} & All   & $-1.458\pm0.021^{\mathrm a}$ & $-1.451\pm0.022$ & $-1.401\pm0.030$ & $-1.412\pm0.025$ \\
         {\it V} & Short & $-1.40\pm0.03$               & $-1.295\pm0.036$ & $-1.284\pm0.041$ & $-1.295\pm0.038$ \\
         {\it V} & Long  & $-1.75\pm0.20$               & $-1.594\pm0.135$ & $-1.795\pm0.298$ & $-1.523\pm0.198$ \\
         {\it I} & All   & $-1.942\pm0.014^{\mathrm a}$ & $-1.896\pm0.015$ & $-1.889\pm0.019$ & $-1.890\pm0.017$ \\
         {\it I} & Short & $-1.90\pm0.02$               & $-1.806\pm0.024$ & $-1.813\pm0.027$ & $-1.817\pm0.026$ \\
         {\it I} & Long  & $-2.09\pm0.15$               & $-2.044\pm0.111$ & $-2.109\pm0.212$ & $-1.909\pm0.138$ \\
         {\it B} & All   & $\cdots$                     & $-1.160\pm0.029$ & $\cdots$ & $\cdots$ \\
         {\it B} & Short & $-1.18\pm0.05$               & $-0.955\pm0.049$ & $\cdots$ & $\cdots$ \\
         {\it B} & Long  & $-1.65\pm0.74^{\mathrm b}$   & $-1.364\pm0.177$ & $\cdots$ & $\cdots$ \\
         \hline
       \end{tabular}
       \begin{list}{}{}
       \item   $^{\mathrm a}$ Since \citet{tam02} does not give the results from the fit to all Cepheids, we adopted the zero-points from \citet{uda99a} because the same dataset is used in both papers.
       \item   $^{\mathrm b}$ The number of long period Cepheids in {\it B}-band is $\sim13$, hence the error is larger than the others. 
       \end{list} 
     \end{table*}

     The existence of two LMC PL relations suggests that in future distance scale studies, the appropriate LMC PL relation may need to be applied to the long and short period Cepheids, respectively \citep[see, e.g.,][]{kan03}. However, all of the previous applications of the LMC PL relation were based on the linear version (in a sense, it is an approximation of two PL relations). Hence an immediate question that arises is: how does the existence of two LMC PL relations affect previous studies? (A similar question was also asked by \citealt{fea03}.) In this paper we show that if the Cepheid distance to a galaxy is derived using the Wesenheit function (or the equivalent $\mu_0$ in equation [4]), for example in the $H_0$ Key Project \citep[][and references therein]{fre01} or in the Araucaria Project \citep{gie04,pie04}, then the results might not be affected (see Section 2.3 for details). This is because the Wesenheit function is linear, even though the PL and PC relations that make up the Wesenheit function are not, as shown in Section 2.2. Given that the LMC PL relation is not linear, and the Wesenheit function has been applied in many papers, we would like to examine the effect of broken LMC PL and PC relations on the application of the Wesenheit function to the distance scale. Again, this will be examined in the optical bands because the {\it V}- and {\it I}-band LMC PL relations have been frequently applied in the literature \citep[as in, e.g.,][]{fre01}.

     In addition to the non-linearity of the LMC PL relation, there are some recent studies also suggesting that the Cepheid PL relation is not universal, i.e., the Galactic PL relation is steeper than the LMC counterparts \citep{tam03,fou03,kan03,nge04,sto04}. It is possible that the Wesenheit function may or may not depend on metallicity: Some preliminary studies for both viewpoints can be found in the literature \citep[see, e.g.,][]{mof86,cap00,bar01,pie04,sto04}. However the detailed study of the metallicity dependency of the Wesenheit function is beyond the scope of this paper. This paper only studies the linearity of the Wesenheit Function for the LMC Cepheids.

\section{The Wesenheit Function and its Application in Distance Scale}
     
\subsection{Definition}

     The Wesenheit function \citep[e.g., see][]{fre88,fre91,gro00,mad76,mad82,mad91,mof86,tan97,uda99a} is defined as $W=\mathrm{magnitude} - R\times \mathrm{colour}$, where $R$ is the ratio of total-to-selective absorption that has to be adopted. Note that the definition of $W$ depends on the adopted $R$ which may be different in different environments. A few variations of $W$ used in the literature with different combinations of magnitudes and colours are:

     \begin{eqnarray}
       W^{BV}_V & = & V - R(B-V),\ R=A_V / E(B-V), \\
       W^{VI}_V & = & V - R(V-I),\ R=A_V / E(V-I), \\
       W^{VI}_I & = & I - R(V-I),\ R=R_I, 
     \end{eqnarray}
       
     \ni where $B$, $V$ \& $I$ denote the (intensity) mean magnitudes. Similar definitions of $W$ in near infrared bands can be found, for example, in \citet{per04}. The other definition of $W$, as given by \citet{vdb75}, is different than the one given in equation (1). The van den Bergh version of $W$ replaces $R$ by the slope of the constant-period line in the colour-magnitude diagram (CMD). \citet{mad91} (also in \citealt{mof86}) have pointed out some problems with the van den Bergh version and the advantage of using $R$ in the definition of $W$. The biggest advantage of using $W$ is that it is reddening-free (see, for example, \citealt{mad91}), i.e. $W =  V - R(B-V) = V_0 - R(B-V)_0 \equiv W_0$, where $V_o$ and $(B-V)_0$ denote the intrinsic visual magnitude and colour, as the effect of interstellar extinction on the observed magnitude and colour cancel out for a star (not only for Cepheids). Another advantage of using $W$ is that the scatter in the $W$-$\log(P)$ plot is reduced \citep{mad82,mad91,bov97,tan97,tan99a,uda99a}, as compared to the scatter in the {\it V}- or {\it I}-band PL relations (see Figure \ref{fig1}). The remaining scatter is due to the combination of photometric errors and the finite width of the instability strip \citep[for example, see ][]{bro80,mad91,bov97,gie98}. Furthermore, an equivalent definition of $W$ with the combination of absolute magnitude and colour can be formulated, as $W_{\mathrm{M}}=M_V-R(M_V-M_I)$, then the distance modulus can be obtained, i.e. $\mu_W=W-W_{\mathrm{M}}$. It is straight forward to show that the following equation:

     \begin{eqnarray}
       \mu_0 & = & \mu_V - R(\mu_V-\mu_I) 
     \end{eqnarray}

     \ni is equivalent to $\mu_W$, where equation (4) is frequently applied in determining the extra-galactic Cepheid distances \citep[see, for example,][]{all04,fre01,kan03,sah01,tan99c}. Therefore, using equation (4) to obtain the distance modulus is equivalent to obtaining the distance modulus by fitting the $W$-$\log(P)$ plane with the empirical $W_{\mathrm{M}}$-$\log(P)$ relation.

\subsection{Testing the Linearity of the Wesenheit Function}
     
     Both the Wesenheit function and equation (4) can be written as a combination of {\it V}- and {\it I}-band PL relations, and they are adopted from the LMC PL relations to derive distances. However, as we mentioned in the Introduction, the LMC PL relations are not linear, hence the applicability of the Wesenheit function and equation (4) is immediately in question. In this sub-section we would like to test the linearity of the Wesenheit function as follows. 

     The PL relation in bandpass $\lambda$ can be written as: $M_{\lambda}=\alpha_{\lambda}^X+\beta_{\lambda}^X\log(P)$. The superscript $X$ denotes the adopted period range, which is either for short ($S$, $\log[P]<1.0$), long ($L$, $\log[P]>1.0$) or all ($A$, short$+$long) periods. Then the $V$-$(V-I)$ Wesenheit function becomes (similar expressions can be derived for other magnitude-colour combinations):

     \begin{eqnarray}
       W^X = (1-R)\alpha^X_V+R\alpha^X_I + [(1-R)\beta^X_V+R\beta^X_I]\log(P),
     \end{eqnarray}

     \ni The linearity of $W$ demands that: 

     \begin{eqnarray}
       (1-R)\beta^S_V + R\beta^S_I & = & (1-R)\beta^L_V + R\beta^L_I \ \ \mathrm{(for\ \ slope)},\\
       (1-R)\alpha^S_V+R\alpha^S_I & = & (1-R)\alpha^L_V+R\alpha^L_I\ \ \mathrm{(for\ \ ZP)}.
     \end{eqnarray}
     
%************************************************
%  FIGURE 2
%************************************************

	\begin{figure}
	\hbox{\hspace{0.1cm}\epsfxsize=8.0cm \epsfbox{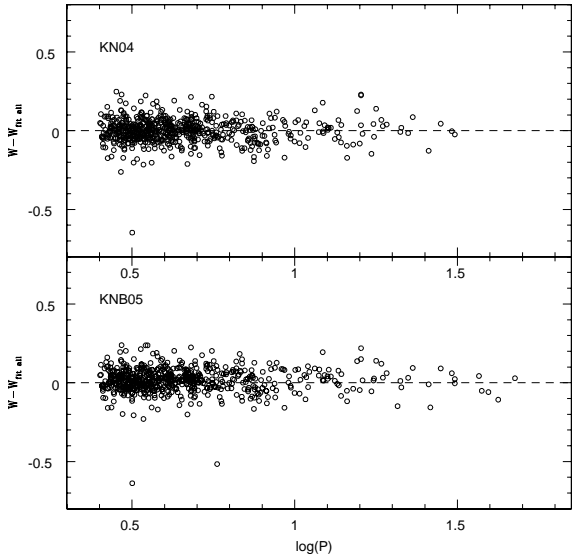}}
	\caption{The residuals of $W$ from Figure \ref{fig1}. The dashed line indicates the zero residuals. The residuals are scattered around the dashed lines, indicating that the Wesenheit function is linear. }
	\label{fig1c}
	\end{figure}

     \ni By using the slopes in Table \ref{tab1}, we can calculate the values of the left-hand side and right-hand side in equation (6), as well as the slope for $W^A$ by using the unbroken PL relation with equation (5). The results are summarized in Table \ref{tab3} with different magnitude-colour combinations. The adopted values for $R$ in these combinations are: $R=A_V/E(B-V)=3.24$ \citep{uda99b}; $R=A_V/E(V-I)=2.45$ \citep{fre01,tan99b}; and $R=R_I=1.55$ \citep{uda99a}. The errors in Table \ref{tab3} are estimated with the standard formula for propagation of errors, i.e. $\sigma^2=(1-R)^2\sigma^2_{\beta_V}+R^2\sigma^2_{\beta_I}$. The same is done for the ZP in Table \ref{tab4}.      

     It can be seen immediately from Table \ref{tab3} \& \ref{tab4} that the short period Wesenheit function is consistent with the long period Wesenheit function, as demanded by equation (6) \& (7). Therefore, the Wesenheit function can be regarded as a linear function of $\log(P)$. Furthermore, the short and long period Wesenheit functions are also consistent with the Wesenheit function obtained from using all Cepheids in the LMC or the linear, unbroken PL relation. Note that the value of $\sim-3.3$ for the slope of the Wesenheit function ($\beta_W$) with $I$-$(V-I)$ combination also agrees with the values given in \citet[][$\beta_W=-3.277\pm0.014$]{uda99a} or in \citet[][$\beta_W=-3.300\pm0.011$]{uda00}. The linearity of the Wesenheit function can be immediately seen from the top panels of Figure \ref{fig1}. The residuals of the $W$ from the fitted regressions in Figure \ref{fig1} are also plotted out as function of period in Figure \ref{fig1c}. If the Wesenheit function is non-linear and can be broken into long and short period Wesenheit functions, as in the LMC PL or PC relations, then the residual plots are expected to show a trend for the long period Cepheids (as in the figure 4 from \citealt{kan04} for the PC relation). However, there is no obvious trend of the residuals seen from Figure \ref{fig1c}. This further supports the linearity of the Wesenheit function.

%************************************
%  TABLE 3 & 4
%************************************

     \begin{table}
       \centering
       \caption{Comparison of the slopes for $W$ obtained using the Cepheids with short, long and all periods.}
       \label{tab3}
       \begin{tabular}{lccc} \hline
         Dataset & Short   & Long   & All \\ 
         \hline \hline
         \multicolumn{4}{c}{$V$-$(V-I)$ Combination, $R=2.45$.}
         \\ \hline
         TR02  & $-3.276\pm0.103$ & $-3.313\pm0.403$ & $-3.255\pm0.068$ \\
         STR04 & $-3.296\pm0.124$ & $-3.192\pm0.253$ & $-3.307\pm0.064$ \\
         KN04  & $-3.311\pm0.141$ & $-3.298\pm0.571$ & $-3.283\pm0.093$ \\
         KNB05 & $-3.312\pm0.133$ & $-3.382\pm0.360$ & $-3.297\pm0.079$ \\
         \hline 
         \multicolumn{4}{c}{$I$-$(V-I)$ Combination, $R=1.55$.}
         \\ \hline
         TR02  & $-3.293\pm0.109$ & $-3.347\pm0.423$ & $-3.275\pm0.072$ \\
         STR04 & $-3.310\pm0.130$ & $-3.217\pm0.266$ & $-3.332\pm0.067$ \\
         KN04  & $-3.325\pm0.149$ & $-3.337\pm0.601$ & $-3.304\pm0.098$ \\
         KNB05 & $-3.327\pm0.140$ & $-3.414\pm0.379$ & $-3.320\pm0.083$ \\
         \hline 
         \multicolumn{4}{c}{$V$-$(B-V)$ Combination, $R=3.24$.}
         \\ \hline
         TR02  & $-4.286\pm0.335$ & $-4.392\pm2.134$ & $\cdots$ \\
         STR04 & $-3.870\pm0.344$ & $-3.915\pm0.613$ & $-3.875\pm0.169$ \\
         \hline 
       \end{tabular}
     \end{table}

     \begin{table}
       \centering
       \caption{Same as Table \ref{tab3}, but for the zero-points.}
       \label{tab4}
       \begin{tabular}{lccc} \hline
         Dataset & Short   & Long   & All \\ 
         \hline \hline
         \multicolumn{4}{c}{$V$-$(V-I)$ Combination, $R=2.45$.}
         \\ \hline
         TR02  & $-2.625\pm0.066$ & $-2.583\pm0.468$ & $-2.644\pm0.046$ \\
         STR04 & $-2.547\pm0.079$ & $-2.696\pm0.335$ & $-2.541\pm0.049$ \\
         KN04  & $-2.580\pm0.089$ & $-2.564\pm0.676$ & $-2.597\pm0.064$ \\
         KNB05 & $-2.574\pm0.084$ & $-2.469\pm0.444$ & $-2.583\pm0.055$ \\
         \hline 
         \multicolumn{4}{c}{$I$-$(V-I)$ Combination, $R=1.55$.}
         \\ \hline
         TR02  & $-2.675\pm0.069$ & $-2.617\pm0.492$ & $-2.692\pm0.048$ \\
         STR04 & $-2.598\pm0.083$ & $-2.741\pm0.352$ & $-2.586\pm0.051$ \\
         KN04  & $-2.633\pm0.094$ & $-2.596\pm0.711$ & $-2.645\pm0.067$ \\
         KNB05 & $-2.626\pm0.089$ & $-2.507\pm0.467$ & $-2.631\pm0.058$ \\
         \hline 
         \multicolumn{4}{c}{$V$-$(B-V)$ Combination, $R=3.30$.}
         \\ \hline
         TR02  & $-2.113\pm0.206$ & $-2.074\pm2.543$ & $\cdots$ \\
         STR04 & $-2.397\pm0.220$ & $-2.339\pm0.810$ & $2.394\pm0.132$ \\
         \hline 
       \end{tabular}
     \end{table}

     A better and more sophisticated test of the linearity is using the $F$-test \citep{wei80,kan04}. The null hypothesis in our $F$-test is that a single linear regression is sufficient, while the alternate hypothesis is that two linear regressions are needed to describe the data. These regressions can be obtained with the standard least squares regression method. The setup and the formalism for the $F$-test is given in \citet{kan04}. By calculating the $F$ value with $N$ data points (note that $F$ is a function of $N$), we can obtain the probability, $p(F)$, under the null hypothesis, for the significance of the $F$ value. In general, the null hypothesis can be rejected if $F$ is large (e.g., $F>4$) or $p(F)$ is small. For example, $p(F)<0.05$ indicates that the null hypothesis can be rejected at $2\sigma$ level. By using the 634 Cepheids as given in \citet{kan04}, we found:

     \begin{description}
     \item  $F=1.473$, $p(F)=0.230$ for $W=V-2.45(V-I)$, 
     \item  $F=0.835$, $p(F)=0.432$ for $W=I-1.55(V-I)$. 
     \end{description}

     \ni Similarly, for the 636 Cepheids in \citet{kan05} the results from the $F$-test are:
     
     \begin{description}
     \item  $F=1.840$, $p(F)=0.160$ for $W=V-2.45(V-I)$,
     \item  $F=1.753$, $p(F)=0.174$ for $W=I-1.55(V-I)$. 
     \end{description}

     \ni Plots of the $W=V-2.45(V-I)$ as function of period for these two datasets are shown in the upper panels of Figure \ref{fig1}. The relatively small values of $F$ and the large values of $p(F)$ indicate that the null hypothesis cannot be rejected. Hence the results from the $F$-test strongly suggest that the Wesenheit function is linear. The results {\it with} the extinction corrections (as given in \citealt{uda99b}) or the removal of the obvious outliers in upper panels of Figure \ref{fig1} and in Figure \ref{fig1c} are essentially the same. 

     By combining the results obtained from this sub-section, we found that, statistically, the Wesenheit function for the LMC Cepheids is indeed linear, although the LMC PL and PC relations are not. Figure \ref{fig1} shows a comparison between the linear Wesenheit function and the broken {\it V}-band PL relation obtained from using two datasets. Therefore, for the LMC Cepheids, we conclude that:

     \begin{description}
     \item {\bf Corollary A:} The Wesenheit function is statistically linear, and the Wesenheit functions for short ($P<10$days), long and all (short+long) period Cepheids are approximately the same, i.e., 
     \end{description}
     \begin{eqnarray}
       W^S \ \ \sim \ \ W^L \ \ \sim \ \ W^A.
     \end{eqnarray}

     \ni The fundamental reason that the Wesenheit function is linear is because the PC relation, e.g. $(V-I)=a+b\log(P)$, is also broken for the LMC Cepheids, in addition to the broken PL relations. The non-linearities for both of the PL and PC relations cancel out when the Wesenheit function is formulated.
     
     In the next subsection, the result from the corollary A (equation [8]) are applied to the distance scale measurements, as well as investigate the accuracy of using the linear LMC PL relation vs. the broken LMC PL relation.
     
\subsection{Distance Scale Application}

     For an ensemble of Cepheids in a target galaxy, with $N$ of them being used in determining the distance, then the distance modulus for the $i^{th}$ Cepheid in bandpass $\lambda$ (usually {\it V} and {\it I}) is $\mu^i_{\lambda}=m^i_{\lambda}-\beta_{\lambda}\log(P_i)-\alpha_{\lambda}$. The distance modulus in bandpass $\lambda$ for these Cepheids can be obtained by taking the unweighted mean to individual Cepheids, i.e. $\overline{\mu_{\lambda}}=\frac{1}{N}\sum_{i=1}^{N} \mu^i_{\lambda}$. The reddening-free distance modulus for the $i^{th}$ Cepheid can be calculated with equation (4), and it is straight forward to show that:

     \begin{eqnarray}
       \overline{\mu_0} & = & \frac{1}{N}\sum_{i=1}^{N} \mu^i_0 \\
                        & = & \overline{\mu_V}-R(\overline{\mu_V}-\overline{\mu_I}),
     \end{eqnarray}

     \ni where $\mu^i_0=\mu^i_V-R(\mu^i_V-\mu^i_I)$. This procedure \citep[see, e.g.,][]{fre01,leo03,kan03,tan97}, i.e. calculating the distance modulus for individual Cepheids and taking the unweighted average to be the final distance modulus to that target galaxy, is equivalent to fitting the {\it V}- and {\it I}-band PL relations to the data and obtaining the $\mu_V$ and $\mu_I$ that apply to equation (4). If the PL relation is linear, then the single, unbroken PL relation (in both {\it V}- and {\it I}-band) can be adopted to fit to all Cepheids to obtain the distance modulus ($\mu_0^A$). However, if the PL relation is non-linear, as in the case of LMC Cepheids, then the long period PL relation should be applied to the long period Cepheids, and similarly for the short period Cepheids. For example, attempts to use the broken LMC PL relations to calibrate the extra-galactic Cepheid distances can be found in \citet{kan03,leo03,thi03,thi04}.

     As mentioned before, $\mu_0$ from equation (4) is equivalent to $\mu_W=W-W_{\mathrm{M}}$, i.e., $\mu_0=\mu_W$. Since the Wesenheit function (both $W$ and $W_{\mathrm{M}}$) for LMC Cepheids is linear (from corollary A or equation [8]), then the $\mu_0$ obtained from using either the broken LMC PL relation (for long and short period Cepheids respectively) or the linear LMC PL relation (for all Cepheids) is expected to agree well with each other. In other words, the difference in the distance modulus ($\Delta \mu_0$) when using either the linear or the broken PL relation should be small. For example, the difference in distance modulus ($\Delta \mu_0=\mu_0^A-\mu_0^{L}$) when using either the linear or the broken long period LMC PL relation can be quantitatively estimated if the mean log-period ($\overline{\log[P]_L}$) of the long period Cepheids in the target galaxy is known. For the distance modulus obtained with equation (4) \& (9), $\Delta \mu_0$ is expressed as \citep{kan03}:

     \begin{eqnarray}
       \Delta \mu_0 & = & -\Delta \beta_W \overline{\log(P)_L} - \Delta \alpha_W,
     \end{eqnarray}
     
     \ni where $\Delta \beta_W$ and $\Delta \alpha_W$ are the differences in slopes and ZPs for the linear and the broken long period Wesenheit function. These coefficients can be calculated using the slopes and ZPs for the $V$-$(V-I)$ Wesenheit function as given in Table \ref{tab3} \& \ref{tab4}, respectively. If $\overline{\log(P)_L}\sim1.4$, then the value of $\Delta \mu_0$ estimated from equation (11) is around $\pm0.03$mag., corresponding to $\sim 1.5$\% change in distance (in Mpc). Furthermore, for $1.0<\overline{\log(P)_L}<2.0$, equation (11) suggests that the maximum $\Delta \mu_0$ when using the linear or the broken long period PL relation from Table \ref{tab1} \& \ref{tab2} is $\sim \pm0.07$mag. A similar estimation can be done for the short period Cepheids (with $0.0<\overline{\log[P]_S}<1.0$) with maximum $\Delta \mu_0$ of $\sim \pm0.02$mag. Even though the value of $\Delta \mu_0$ is small, it has a {\it systematic} effect on the distance scale because of the different PL relations used. 

     We give two examples to illustrate the above discussion further:

     \begin{enumerate}
     \item \citet{kan03} calculated the distance moduli to 25 {\it HST} observed galaxies with the linear and the broken long period LMC PL relations (as $\sim98$\% of the Cepheids detected in these galaxies have period longer that 10 days). By comparing the distance moduli derived from the linear PL relation and the broken long period PL relation, the unweighted average of $\Delta \mu_0$ in these 25 galaxies is $-0.021\pm0.002$mag., in the sense that the linear PL relation {\it systematically} produces slightly smaller distance moduli than the broken long period PL relation. Since the mean log-period for most of these galaxies is $\sim1.4$ \citep{kan03}, then the difference of $-0.02$mag. is expected and agrees well with the above discussion. 
     
     \item To compare the distance moduli obtained from using the linear and the broken PL relations, we use the Cepheids in IC 4182 \citep[with the data from][]{gib00} as an example, because this galaxy contains roughly equal number of short ($N=13$) and long ($N=15$) period Cepheids. The distance moduli from using the linear, unbroken PL relation with all 28 Cepheids ($\mu_0^A$) are compared with the mean distance moduli ($\overline{\mu_0}$, from equation [9]) obtained from applying the broken long and short period PL relation to the individual long and short period Cepheids, respectively. The PL relations used, include the linear and the broken PL relations, are taken from Table \ref{tab1} \& \ref{tab2}, and the results are presented in Table \ref{tab5}. From this table, $\mu_0^A$ shows a good agreement to $\overline{\mu_0}$, with a difference of $\sim0.01$mag. or smaller\footnote{Note that the median values of $\sim 28.25$mag. for $\mu_0^A$ and $\overline{\mu_0}$ in Table \ref{tab5} agree well to the distance modulus obtained from the TRGB (tip of red giant branch) method as given in \citet[][$\mu_{TRGB}=28.25\pm0.06$mag.]{sak04}.}. 
     \end{enumerate}

     In short, when using the LMC PL relations to derive the Cepheid distances, we conclude that:
     
     \begin{description}
     \item {\bf Corollary B:} The distance moduli are approximately the same when using either the linear PL relation (to all Cepheids) or the means with broken PL relation (to long and short period Cepheids respectively), i.e., 
     \end{description}
     \begin{eqnarray}
       \mu_0^A \ \ \ \sim \ \ \ \overline{\mu_0}.
     \end{eqnarray}
     
     \ni Both of the approaches will give similar and consistent distance moduli. Recall that the linear PL relation is the approximation of the broken PL relation, and the accuracy of using the linear PL relation is roughly $\pm0.03$mag. Therefore, researchers have the freedom to use either the linear LMC PL relation as an approximation or using the correct broken LMC PL relation in deriving the Cepheid distances.

% ************************************
% TABLE 5 
% ************************************

     \begin{table}
       \centering
       \caption{The distance modulus ($\mu_0$) to IC 4182.}
       \label{tab5}
       \begin{tabular}{lcc} \hline
         LMC PL relation & $\mu_0^A$ & $\overline{\mu_0}$ \\  
         \hline \hline
         TR02  & $28.267\pm0.053$ & $28.277\pm0.053$ \\
         SRT04 & $28.222\pm0.053$ & $28.217\pm0.053$ \\ 
         KN04  & $28.250\pm0.053$ & $28.247\pm0.053$ \\ 
         KNB05 & $28.253\pm0.053$ & $28.257\pm0.054$ \\ 
         \hline
       \end{tabular}
     \end{table}

\subsubsection{A Note on the Random Errors for $\mu_0$ from Equation (4)}
     
     By definition, the random error for $\overline{\mu}=\frac{1}{N}\sum_{i=1}^{N} \mu^i$, due to purely statistical fluctuations, is: 
     
     \begin{eqnarray}
       \sigma^2 & = & \frac{1}{N(N-1)}\sum_{i=1}^{N} (\mu^i-\overline{\mu})^2. 
     \end{eqnarray}

     \ni This equation holds for the $\overline{\mu_{\lambda}}$ and $\overline{\mu_0}$ (from equation [9]). We can expand the expression for $\sigma^2_0$ by substituting the expression for $\mu^i_0$ and $\overline{\mu_0}$ to equation (13), then:

  \begin{eqnarray}
    \sigma^2_0 & = & (1-R)^2\sigma^2_V+R^2\sigma^2_I - {\mathrm CORR}_{VI},
  \end{eqnarray}

  \ni where $\sigma^2_V$ and $\sigma^2_I$ are given by equation (13), and $\mathrm{CORR}_{VI}=\frac{2(R-1)R}{N(N-1)}\sum_{i=1}^{N} (\mu^i_V-\overline{\mu_V}) (\mu^i_I-\overline{\mu_I})$, a term for the correlated residuals from both bands \citep{tan97,fre01}. Note that the use of equation (13) to estimate the random errors for the Cepheid distance modulus has been practiced in the literature \citep[e.g., see][]{fre01,kan03}.

  On the other hand, if we apply the standard equation for the propagation of errors (POE) to equation (4) or (10), by ignoring the correlation term, then we have $\sigma^2_0(POE)=(1-R)^2\sigma^2_V+R^2\sigma^2_I$. By comparing this expression to equation (14), we obtain: $\sigma^2_0(POE) = \sigma^2_0 + \mathrm{CORR}_{VI}$. One can immediately see that $\sigma^2_0(POE)$ is greater than $\sigma^2_0$ since the term $\mathrm{CORR}_{VI}$ is mostly likely to be positive. This is because (a) $R>1$ from the extinction curve; and (b) $\sum (\mu^i_V-\overline{\mu_V}) (\mu^i_I-\overline{\mu_I})$ is mostly likely to be positive. The second condition is due to the existence of the period-luminosity-colour relation: if the {\it V}-band magnitude for a Cepheid is above (or below) the ridge-line of the {\it V}-band PL relation, then the corresponding {\it I}-band magnitude will also be above (or below) the ridge-line of the {\it I}-band PL relation. The result is that if $(\mu^i_V-\overline{\mu_V})$ is positive (or negative), then $(\mu^i_I-\overline{\mu_I})$ is also positive (or negative), and hence the product of these two is positive. Further, if the extinction and/or the correlated errors (from measurement) make $(\mu^i_V-\overline{\mu_V})$ to be positive/negative, then $(\mu^i_I-\overline{\mu_I})$ is also going to be positive/negative, and again the product will be positive. Errors estimated from error propagation with equation (10), i.e. $\sigma^2_0(POE)$, will ignore the $CORR_{VI}$ term and hence resulted a larger random error than the random error estimated from equation (13). Note that all the errors discussed here, the $\sigma$, are random errors only, which do not include the systematic errors such as the errors arise from the calibrations, the width of the instability strip, and others \citep[see, e.g., the discussion by][]{sah00}.

\section{Conclusion}

     Due to the recent discovery of the non-linearity for the LMC PL and PC relations, \citet{fea03} has asked the following critical question:
     
     \begin{quote}
       ``$\cdots$ is there a significant slope difference between short and long ($>\sim10$days) Cepheids that would seriously affect the calibration and use of PL [and PC] relation?''
     \end{quote}
     
     \ni From this study, we showed that this problem can be remedied with the application of the Wesenheit function in distance scale studies. This is because the Wesenheit function for the LMC Cepheids is linear (corollary A), as shown in Section 2.2, although the LMC PL and PC relations are not. Therefore, the Cepheid distances obtained with the Wesenheit function or the equivalent $\mu_0$ would not be affected with the recent finding of the broken PL and PC relations. We also found that the typical difference in distance modulus from using the linear or the broken PL relations is about $\pm0.03$mag. Hence, researchers can choose to apply either the linear or the broken LMC PL relations to obtain the Cepheid distances, without worrying that these two approaches will give inconsistent results, as both approaches are equally applicable in deriving the Cepheid distance (corollary B). Therefore, the broken PL relation found in the LMC Cepheids will not seriously affect the previous applications of the linear LMC PL relation in distance scale studies because the effect is minimal. The question raised by \citet{fea03} is essentially answered.

%********************************************************
%  acknowledgments
%********************************************************

\section*{acknowledgements}

We would like to thank the useful discussion with W. Gieren, D. Leonard, S. Nikolaev and G. Tammann. We would also like to thank the anonymous referee for useful suggestions.

%***************************************************
%  REFERENCE 
%***************************************************

\end{document}